\documentclass[preprint,byrevtex,showpacs, prb]{revtex4}
\usepackage{graphicx}
\usepackage{dcolumn}
\usepackage{bm}
\usepackage{epsfig}
\usepackage{times}
\usepackage{amsmath, amsthm}

\begin{document}
\title{Dispersion and transitions of dipolar plasmon modes
in graded plasmonic waveguides}

\author{J. J. Xiao\footnote{Electronic address: jjxiao@phy.cuhk.edu.hk}}
\affiliation {Department of Physics, The Chinese University of Hong Kong,
Shatin, New Territories, Hong Kong, China}
\author{K. Yakubo}
\affiliation{Division of Applied Physics, Graduate School of Engineering,
Hokkaido University, N13-W8, Sapporo 060-8628, Japan}
\author{K. W. Yu}
\affiliation{Department of Physics and Institute of Theoretical Physics, The
Chinese University of Hong Kong, Shatin, New Territories, Hong Kong, China}

\begin{abstract}
Coupled plasmon modes are studied in graded plasmonic waveguides, which are
periodic chains of metallic nanoparticles embedded in a host with gradually
varying refractive indices. We identify three types of localized modes called
``light", ``heavy", and ``light-heavy" plasmonic gradons outside the
passband, according to various localization. We also demonstrate different
transitions among extended and localized modes when the interparticle
separation $d$ is smaller than a critical $d_c$, whereas the three types of
localized modes occur for $d>d_c$, with no extended modes. The transitions
can be explained with phase diagrams constructed for the lossless metallic
systems.
\end{abstract}
\date{\today}

\maketitle

Recent explorations of surface plasmon (SP) waves sustained by metallic
nanostructures have been successful in nanooptics, see, e.g., Refs.~1-2.
Specifically, guiding and manipulating electromagnetic (EM) energy below the
diffraction limit with SPs have attracted much effort, concentrating on
routing  plasmon signals in periodic noble metal nanoparticle chains.
\cite{Quinten98,Brongersma00,Weber04, Citrin06} In these so called plasmoic
waveguides (PWs), metal nanoparticles can sustain resonant collective
oscillations (plasmons) of their conduction electrons. When driven by an
external light field, these electron oscillations couple to the visible
optical excitation to form SPs bound around the nanoparticles, therefore
possess a lot of novel properties. \cite{Barnes03,
Atwater05,Weber04,Quinten98,Citrin06,Brongersma00} The dispersion relations
of coupled SP modes in these PWs are like those of propagative EM waves in
one dimensional Bragg gratings, \cite{Brongersma00} rather more like electron
dispersions within ``tight-binding" atoms in a solid. The plasmon resonant
bands, which are quasicontinuous for finite system and continuous for
infinite system, are determined by the interparticle couplings and almost
centered by individual particle's site Mie resonances
\cite{Brongersma00,Weber04} which are, crucially dependent on a rich set of
parameters  such as the nanoparticle's size, shape, as well as surrounding
medium. \cite{Atwater05,Krenn05}

In PWs, however, longitudinal confinements of the plasmon excitation should
also be of great interest and have been explored little so far. Here we
propose to use a host with graded refractive indices along the nanoparticle
chain---graded plasmonic waveguide (GPW). This belongs to a subclass of
deterministic nonperiodic media which has recently been revealed to exhibit
interesting spectrum transitions in both elastic and plasmonic systems.
\cite{gradon,Swith06} As compared to the case of homogeneous host (e.g.,
vacuum), apparently a graded host has two consequences: (i) shift of Mie
resonance of individual particle, simply because dielectrically denser host
results in a heavier effective optical electron mass $m^*$ in the metallic
particle;\cite{Brongersma00} (ii) screening effect on the interparticle
couplings. It turns out that GPWs are far more intriguing than expected; very
rich longitudinal (axial) localization effects are discovered.

We consider a linear chain of $N$ spherical metallic nanoparticles immersed
in a dielectric host. We take lossy Drude form dielectric function
$\epsilon(\omega)=1-\omega_p^2/\omega(\omega+i \Gamma)$ for the nanoparticles
while assume that the host has its dielectric constant varying from the
chain's left-hand side to the right-hand side along $x$ axis as
$\epsilon_2(x_n)=\epsilon_{\ell}+cx_n/l$, ($n=1, 2, \cdots,N$), where $x_n$
denotes the position of the $n$-th nanoparticles, $l$ the total length of the
chain, and $c$ the coefficient of dielectric gradient. A point dipole model
is believed sufficient in capturing the fundamental dispersion
characteristics of coupled plasmons in PWs, \cite{Weber04,Brongersma00} we
expect this to be also true in GPWs. The coupled equation for point dipoles
reads
\begin{equation} {\boldsymbol G}(\omega) {\boldsymbol p}=0,
\label{eq:eigenequation}
\end{equation}
where $\boldsymbol{p}$ is the $N$-rowed column vector of 
dipole moment oscillating with the frequency $\omega$, ${G}_{nn}(\omega) =
1/\beta_n(\omega), (n = 1, \dots ,N)$ and ${G}_{nm}(\omega) = -{\tilde
T}_{nm}$, ($m \neq n$). Here
$\beta_n(\omega)=\epsilon_2(x_n)[\epsilon(\omega)-\epsilon_2(x_n)]
/[\epsilon(\omega)+2\epsilon_2(x_n)]$ represents the polarizability of the
$n$-th particle, and ${\tilde T_{nm}}$ denote the EM interaction on the
$n$-th particle due to the $m$-th particle. \cite{Weber04}
Equation~\eqref{eq:eigenequation} can also be regarded as an eigenequation
with normal mode frequency $\omega_{\alpha}$ determined by
$\text{det}\{{\boldsymbol G}(\omega)\}=0$, while $\omega_{\alpha}$ generally
being complex valued. However, it is hard to determine $\omega_{\alpha}$ by
directly solving this equation which is a nonlinear complex transcendental
equation with respect to $\omega$. Alternatively, we employ the inhomogeneous
equation ${\boldsymbol G}(\omega) {\boldsymbol p}=v$ to study the plasmon
excitations, as in
Ref.~5.

\begin{figure*}[htbp]
\includegraphics*[clip,width=0.45\textwidth]{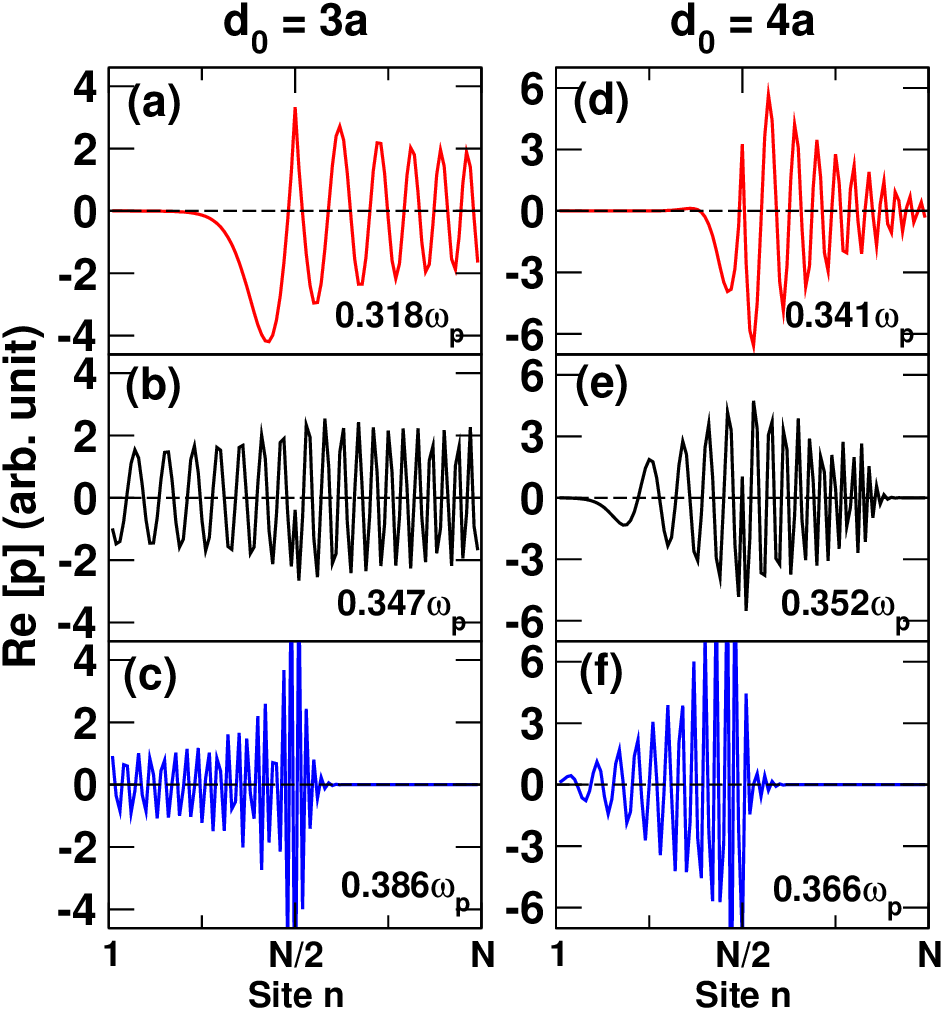}
\caption{/Xiao, Yakubo, and Yu} \label{fig:FOMmodes}
\end{figure*}

Figure~\ref{fig:FOMmodes} shows, for the longitudinal polarization (along the
chain axis $\hat{x}$), the real part of dipole moments obtained via
${\boldsymbol p}={\boldsymbol G}^{-1}(\omega) {\boldsymbol v}$, where
${\boldsymbol G}^{-1}(\omega)$ is evaluated at a real valued frequency
$\omega$ indicated in every panel. \cite{Weber04} In these calculations, we
have set all but the row of $N/2$ in ${\boldsymbol v}$ (a column vector)
equal to zero. The GPW parameters are $\epsilon_{\ell}=3.0$, $c=1.0$,
$N=100$, and interparticle distance $d_0=4a$ ($3a$) for the left (right)
panels, where $a$ is the radii of the nanoparticles. To highlight the graded
effect, only near-field couplings (${\tilde T_{nm}} \propto
1/\left|x_n-x_m\right|^3$) are taken into account. We note that the fully
retarded couplings were shown to be remarkably important in the case of
transverse polarization. \cite{Weber04,Citrin06} However, their role in
modifying the dispersion is a subject of controversy. \cite{Polman06}
These issues become more ambiguous in the presence of loss (radiative or
nonradiative), which needs to be addressed appropriately further. Here we
mainly focus on the longitudinal polarization.\cite{note0} The demonstrated
localizations shall hold even with full couplings
included,\cite{Weber04,Citrin06,Polman06} in view of a recent exploration in
the localized vibrations of optically (long-range) bounded
structures.\cite{Chan06} It is seen in Fig.~\ref{fig:FOMmodes} that there
exist several different solutions which correspond to modes excited by
external periodic ``forces" subject to the central site $n = N/2$ with
different frequencies: (1) at relatively low frequency, modes are excited in
the right-hand side (larger $\epsilon_2$), which implies the externally
injected energy moves to the right [Figs.~\ref{fig:FOMmodes}(a) and
\ref{fig:FOMmodes}(d)]; (2) at intermediate frequency, it propagates to both
sides [Figs.~\ref{fig:FOMmodes}(b) and \ref{fig:FOMmodes}(e)]; (3) while at
high frequency, it tends going to the left [Figs.~\ref{fig:FOMmodes}(c) and
\ref{fig:FOMmodes}(f)]. We can regard these excitation modes as being
localized or extended depending on their
spatial extent. However, if $\Gamma$ is large, 
we can not simply regard the solutions (not shown) in this way. This is
because it is hard to distinguish the localization effect from the loss which
leads to attenuation of the excitation in the medium. \cite{Busch05}


In order to gain more insights into the peculiar and principal
characteristics of GPWs, we therefore focus on weak loss or lossless regime
hereafter.
It is possible to linearize Eq.~\eqref{eq:eigenequation} with respect to
$\omega^2$
\begin{equation}
({\boldsymbol F}-\omega^2{\boldsymbol I}) {\boldsymbol p}=0,
\label{eq:linearized}
\end{equation}
where $\boldsymbol I$ is identity matrix and $F_{nm}=\omega_n^2 -{
G_{nm}}(\omega_n^2)/G^{\prime}_{nn}(\omega_n^2)$. \cite{Swith06}
Here 
the prime indicates derivative with respect to $\omega^2$
and $\omega_n$ is the dipolar Mie resonant frequency of the $n$-th particle.
Equation~\eqref{eq:linearized} represents a good approximation of
Eq.~\eqref{eq:eigenequation} at $\omega \approx \omega_n$.
In considering the case of weak loss, we can write the eigenspectrum \{$
\omega_{\alpha} =\omega^{\prime}_{\alpha} + i\omega^{\prime\prime}_{\alpha},
{{\boldsymbol p}_{\alpha}}\}$ and ${\boldsymbol v} = \sum_{\alpha}
\mu_{\alpha} {{\boldsymbol p}_{\alpha}}$.
If the frequency $\omega$ is
close to a eigenfrequency $\omega_{\alpha_0}$ 
then significant contribution to ${\boldsymbol p}$ comes from a few
eigenmodes whose eigenfrequency is close to $\omega$. 
Let us call the set of modes $S$ that has significant contribution:
$|\omega^{\prime}_{\alpha}-\omega^{\prime}_{\alpha_0}| \ll
\omega^{\prime\prime}_{\alpha}~~ \text{for}~~ {\alpha} \in S.$
Thus, the real part of the response function (e.g., plotted in
Fig.~\ref{fig:FOMmodes}) is
$\text{Re}[ {\boldsymbol p} ] = \sum_{\alpha \in S} \mu_{\alpha} (\omega -
\omega^{\prime}_{\alpha}){{\boldsymbol p}}_{\alpha}/2\omega [(\omega -
\omega^{\prime}_{\alpha})^2 + {\omega^{\prime\prime}_{\alpha}}^2]$.
This means by solving  ${\boldsymbol p}={\boldsymbol G}^{-1}(\omega)
{\boldsymbol v}$ for weak loss in the Drude model, the solutions we get in
Figs.~\ref{fig:FOMmodes}(a)--\ref{fig:FOMmodes}(f) are very close to a single
complex eigenmodes in the GPWs. 

\begin{figure*}[htbp]
\includegraphics*[clip,width=0.45\textwidth]{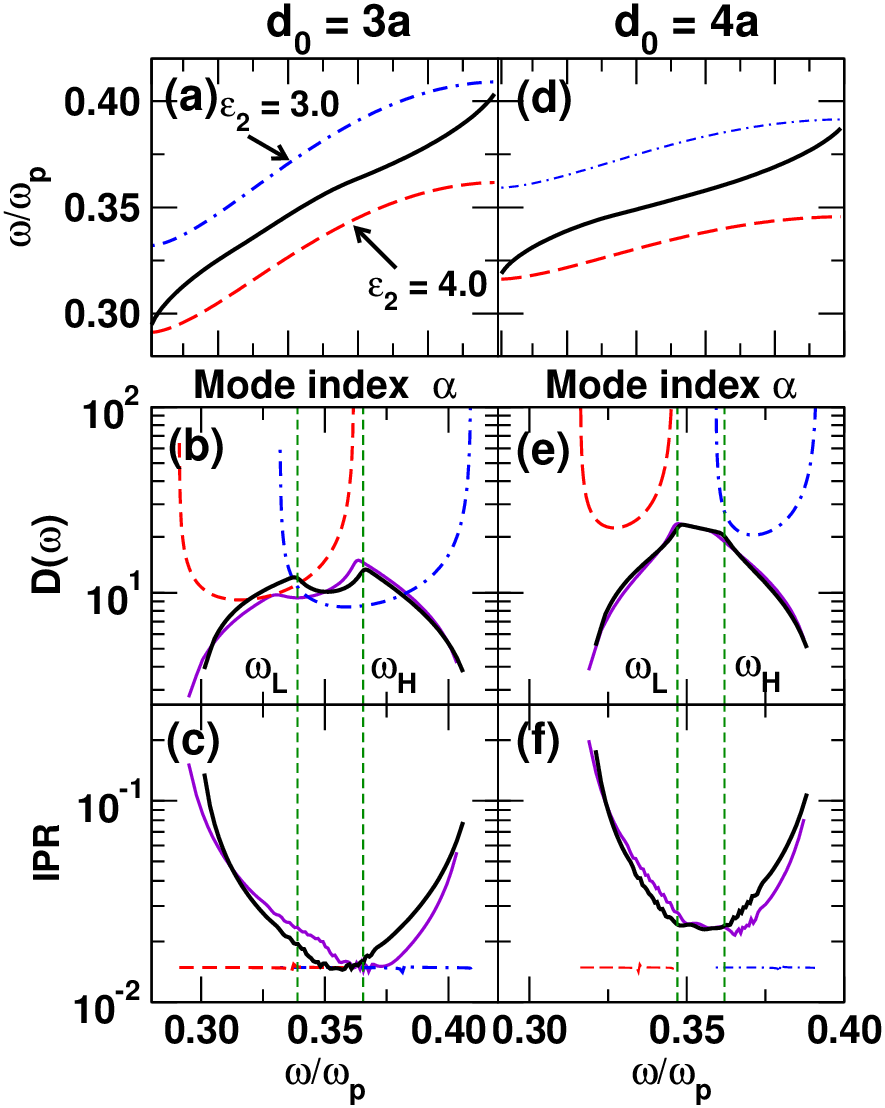}
\caption{/Xiao, Yakubo, and Yu} \label{fig:lightheavy}
\end{figure*}

In the linearized lossless  case, the coupled plasmon dispersion can be
directly obtained by diagonalizing ${\boldsymbol F}$, resulting in real
eigenspectrum \{$\omega_{\alpha}, {\boldsymbol p}_{\alpha}\}$. \cite{Swith06}
The results are shown in Figs.~\ref{fig:lightheavy}(a) and
\ref{fig:lightheavy}(d) (solid lines). Still these are in the near-field
coupling approximation, for the same GPW structures studied in
Fig.~\ref{fig:FOMmodes} except for $\Gamma=0$ now. Also the density of states
(DOS) $D(\omega)=\sum_{\alpha}\delta(\omega-\omega_{\alpha})/N$ are shown in
Figs.~\ref{fig:lightheavy}(b) and \ref{fig:lightheavy}(e). The real valued
eigenmodes ${\boldsymbol p}_{\alpha}$ (not shown) have quite different
spatial extents, much like the excitation patterns in
Fig.~\ref{fig:FOMmodes}. This is also reflected by their inverse
participation ratio\cite{Swith06} (IPR) $\sum_{n=1}^N{\boldsymbol
p}_{\alpha}^4(n)/\left[\sum_{n=1}^N {\boldsymbol p}_{\alpha}^2(n)\right]^2$
shown in Figs.~\ref{fig:lightheavy}(c) and \ref{fig:lightheavy}(f). The IPR
is of the order of $1/N$ if modes are extended over the system with $N$
degrees of freedom, while it becomes larger than $1/N$ for spatially
localized modes. We see that relatively low frequency modes are localized at
the side of larger $\epsilon_2$ while high frequency modes are localized at
the side of smaller $\epsilon_2$.
We name these modes as ``heavy gradons" and ``light gradons", respectively.
More interestingly, coupled plasmon modes at intermediate frequencies between
two transition points $\omega_{\text{L}}(d_0,
c)<\omega<\omega_{\text{H}}(d_0, c)$ are extended \cite{Swith06} when
$d_0=3a$  [corresponding to Fig.~\ref{fig:FOMmodes}(b)], but are still
somehow localized, rather at the central part of the GPW for $d_0=4a$
[corresponding to Fig.~\ref{fig:FOMmodes}(e)]. These center-localized modes
resemble light gradons [corresponding to Figs.~\ref{fig:FOMmodes}(c) and
\ref{fig:FOMmodes}(f)] at the left tail while look like heavy gradons
[corresponding to Figs.~\ref{fig:FOMmodes}(a) and \ref{fig:FOMmodes}(d)] in
the right front, therefore are called ``light-heavy gradons."


Equation~\eqref{eq:linearized} can be mapped into an equivalent chain of
graded coupled harmonic oscillators  with additional on-site potentials.
\cite{gradon} In this case, the vibrating mass
$M_{n}=[1+2\epsilon_{2}(x_{n})]^2/3\lambda\omega_{p}^{2}\epsilon_{2}(x_{n})$,
and the strength of the additional harmonic spring
$U_{n}=M_n\omega_n^2-2K_{0}$,
where $K_{0}=(a/d_{0})^3$ is a force constant between adjacent masses,
$\lambda=2$ ($-1$) for longitudinal (transverse) case. \cite{Swith06}
We then define two characteristic frequencies
$\omega_{c1}(n)=\sqrt{U_n/M_n}$ and $\omega_{c2}(n)=\sqrt{(U_n+4K_0)/M_n}$.
It is easy to notice that $\omega_{c1}(n) \to \omega_n$ and $\omega_{c2}(n)
\to \omega_n$ when $d_0 \to \infty$,
which are as expected because the coupling between the particles vanishes for
infinitely large separation.
We previously shown that when $\omega_{c1}(1)<\omega_{c2}(N)$, one get
extended modes between $\omega_{\text{L}}(d_0,
c)<\omega<\omega_{\text{H}}(d_0, c)$, where $\omega_{\text{L}}(d_0,
c)=\omega_{c1}(1)$ and $\omega_{\text{H}}(d_0, c)=\omega_{c2}(N)$.
\cite{Swith06} However, if $\omega_{c1}(1)>\omega_{c2}(N)$, there is no
extended modes but light-heavy gradons when $\omega_{\text{L}}(d_0,
c)<\omega<\omega_{\text{H}}(d_0, c)$, where $\omega_{\text{L}}(d_0,
c)=\omega_{c2}(N)$ and $\omega_{\text{H}}(d_0, c)=\omega_{c1}(1)$.
Furthermore, there exists a critical point of $d_0=d_c$ when
$\omega_{c1}(1)=\omega_{c2}(N)$, showing only one light-heavy gradon which,
however, is across the whole system like an extended mode. These arguments
work well in view of the fact that both the band boundaries and the
transition frequencies [$\omega_{\text{L}}(d_0, c)$ and
$\omega_{\text{H}}(d_0, c)$] agree with the numerical data: the two
transition frequencies are marked by the vertical dashed lines in
Figs.~\ref{fig:lightheavy}(b)--\ref{fig:lightheavy}(f) where they meet with
the singularities in the DOS curve (thick dark lines).

\begin{figure*}[htbp]
\includegraphics*[scale=0.45]{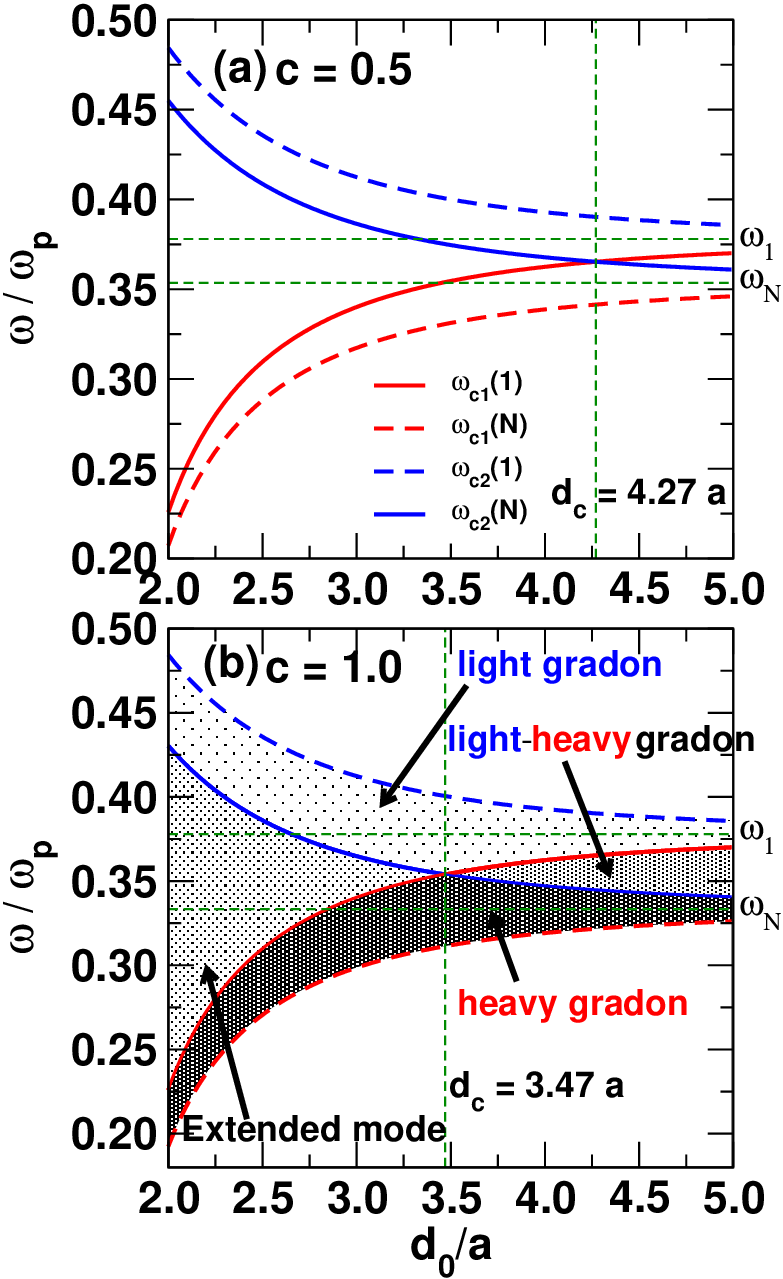}
\caption{/Xiao, Yakubo, and Yu}\label{fig:phase}
\end{figure*}

\begin{figure*}[htbp]
\includegraphics*[clip,width=0.45\textwidth]{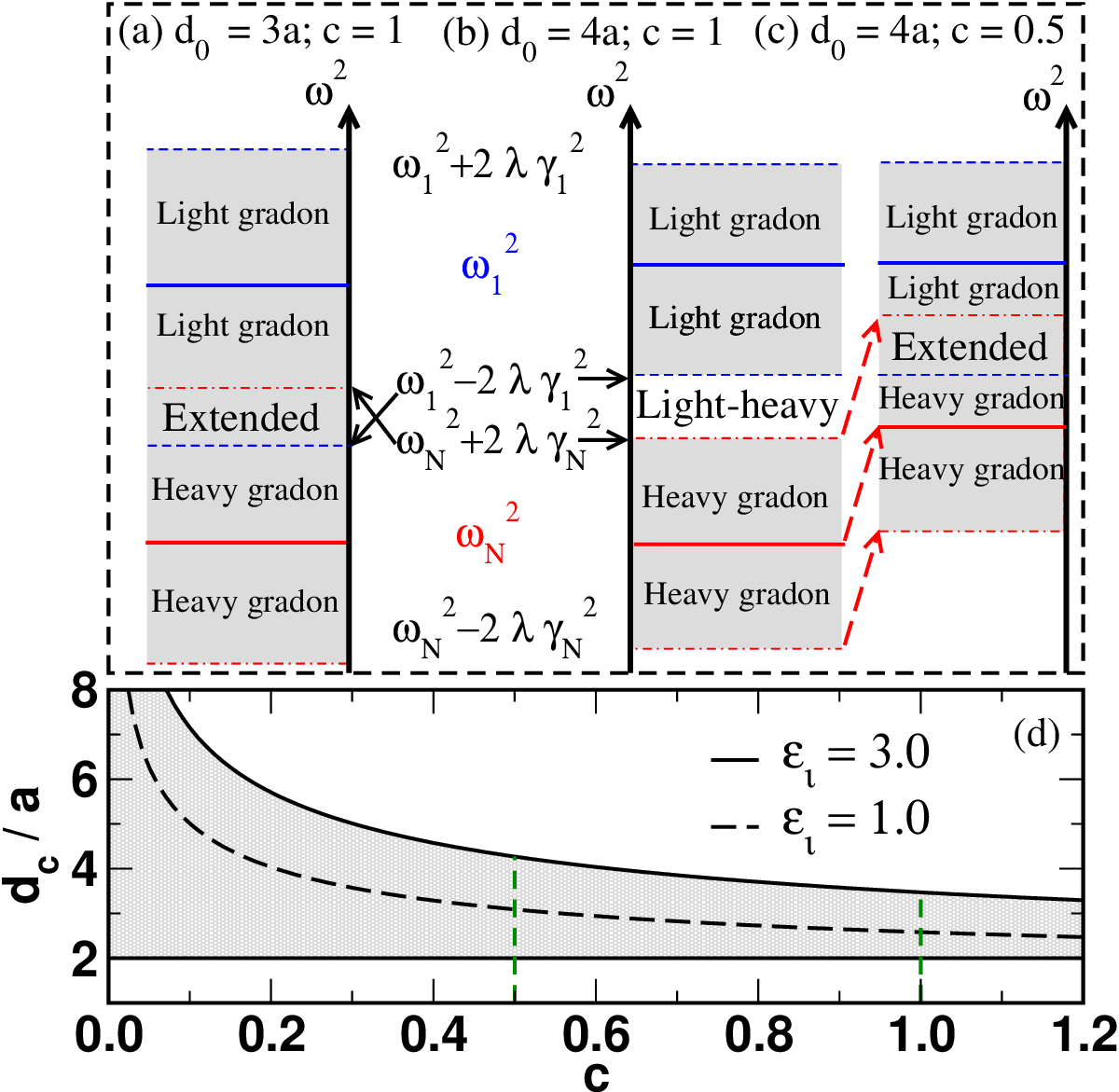}
\caption{/Xiao, Yakubo, and Yu} \label{fig:dc_c}
\end{figure*}

As there are various plasmon modes sustained by GPWs, we construct a phase
diagram as Fig.~\ref{fig:phase}, which shows not only the case of $c=1.0$
[Fig.~\ref{fig:phase}(b)] but also the case of $c=0.5$
[Fig.~\ref{fig:phase}(a)]. Nevertheless, let us focus on
Fig.~\ref{fig:phase}(b) for the discussion. Figure~\ref{fig:phase}(b)
contains four curves of $\omega_{c1}(1)$, $\omega_{c1}(N)$, $\omega_{c1}(1)$,
and $\omega_{c2}(N)$ as functions of $d_0$. As we have mentioned, there
indeed exists a point $d_0=d_c$ (vertical dashed line) when $\omega_{c1}(1)$
and $\omega_{c2}(N)$ cross. Also we notice that
$\omega_{c1}(1)=\omega_{c2}(1) \to \omega_1$ and
$\omega_{c1}(N)=\omega_{c2}(N) \to \omega_N$ as $d_0 \to \infty$, which are
clearly marked by the two horizontal dashed lines of $\omega_1=0.378\omega_p$
and $\omega_{\text{N}}=0.333\omega_p$, respectively. In consistent with the
previous results, we now discuss the four shaded regions partitioned by the
four curves in Fig.~\ref{fig:phase}(b). Specifically, (1) extended mode is
possible only when $\omega_{c1}(1)<\omega<\omega_{c2}(N)$, and when
$d_0<d_c=3.47a$ which corresponds to the left panels in
Figs.~\ref{fig:FOMmodes} and \ref{fig:lightheavy}; (2) light-heavy gradons
emerge when $\omega_{c2}(N)<\omega<\omega_{c1}(1)$ for $d_0>d_c$,
corresponding to the right panels in Figs.~\ref{fig:FOMmodes} and
\ref{fig:lightheavy}; (3) the lower black region indicates a phase of heavy
gradons which have relatively low frequencies; (4) the upper dotted region
indicates a phase of light gradons which have relatively high frequencies.
Similarly, the same phase regions exist in Fig.~\ref{fig:phase}(a), where a
reduced $c=0.5$ results in an increased $d_c=4.27a$, but the discussions on
the four phases through (1)$-$(4) are still applicable. In fact, the critical
$d_c$ is determined by
$\omega_{c1}(1)=\omega_{\text{L}}(d_0,c)=\omega_{\text{H}}(d_0,c)
=\omega_{c2}(N)$ which yields
\begin{eqnarray}
\label{eq:dc_c1}
\frac{1}{1+2\epsilon_2(x_1)}&-&\frac{1}{1+2\epsilon_2(x_\text{N})}= 6\lambda
K_0 \nonumber\\
&\times& \left(\frac{\epsilon_2(x_1)}{[1+2\epsilon_2(x_1)]^2}+
\frac{\epsilon_2(x_{\text{N}})}{[1+2\epsilon_2(x_\text{N})]^2}\right).
\end{eqnarray}
The results from Eq.~\eqref{eq:dc_c1} are shown in Fig.~\ref{fig:dc_c}(d) for
both $\epsilon_{\ell}=3.0$ and $\epsilon_{\ell}=1.0$. This is a guideline for
determination of extended modes. \cite{note} For instance, only the region
below the solid line (shaded region) supports extended modes, for the case of
$\epsilon_{\ell}=3.0$. Note that in this region there exist also light and
heavy gradons, depending on the frequency, whereas the region above the solid
line indicates appearance of localized modes only, orderly in the form of
heavy gradon, light-heavy gradon, and light gradon as frequency increases.

Up to now, we have already interpreted the mode transitions in view of the
equivalent coupled harmonic oscillators of graded masses and on-site
potentials.  From another perspective, we further examine the relationship
between the site Mie resonance of isolated nanoparticle, the resonant band of
an infinite PW, and the resonant band of an infinite GPW with infinitesimal
gradient. In Fig.~\ref{fig:lightheavy} we have already plotted the
corresponding results for the chains in homogeneous host of
$\epsilon_2=\epsilon_{\ell}$ (dash-dotted lines) and
$\epsilon_2=\epsilon_{\ell}+c$ (dashed lines), i.e., results for homogeneous
PWs.
It is known that infinite periodic PWs
have resonant bands around respective Mie resonance $\omega_n$ as
\cite{Weber04,Brongersma00}
\begin{equation}
\label{eq:band} \omega^2=\omega_n^2-2\lambda\gamma_n^2\cos(kd_0)\cosh(\tau
d_0),
\end{equation}
where $k$ is the wave number of the plasmon wave, $\tau$ the attenuation
coefficient.
Here the nearest-neighboring electrodynamic coupling strength $\gamma_n^2=
\gamma_{0n}^2 K_0$, where $\gamma_{0n}$ is host-dependent.
\cite{Brongersma00} The second-order correction due to dissipations is
typically $\cosh(\tau d_0) \sim 1.001$ for silver, \cite{Brongersma00} which
is negligible.
Let us break the GPW into a large number of infinite segments of PWs, each of
which approximated by homogeneous host $\epsilon_2(x_n)$. For each of these
segments we have Eq.~\eqref{eq:band}, which means a series of bands located
at different Mie resonance $\omega_n$. Overlapping of these bands can be used
as guideline to determine mode types, as sketched in Fig.~\ref{fig:dc_c}(a).
In Fig.~\ref{fig:dc_c}(b), increased $d_0$ defies the overlapping of these
bands, therefore extended modes can not show up. For fixed $d_0=4a$,
decreasing the gradient coefficient to $c=0.5$ will overlap all these bands
again, resulting in extended modes [see Fig.~\ref{fig:dc_c}(c)]. 
In this way, the critical interparticle distance $d_c$ is determined by the
condition that the lower band edge of Eq.~\eqref{eq:band} for
$\epsilon_2=\epsilon_2(x_1)$ equals to the upper band edge for
$\epsilon_2=\epsilon_2(x_{\text{N}})$, i.e., $\omega_1^2-2\lambda\gamma_1^2=
\omega_{\text{N}}^2+2\lambda\gamma_{\text{N}}^2$
which simplifies to
\begin{equation}
\label{eq:bandconst}
\frac{1}{1+2\epsilon_2(x_1)}-\frac{1}{1+2\epsilon_2(x_\text{N})}=\lambda K_0
\left(\frac{\gamma_{01}^2}{\omega_p^2}+\frac{\gamma_{0\text{N}}^2}
{\omega_p^2}\right),
\end{equation}
where $\gamma_{0n}$ more explicitly depends on the resonant frequency
$\omega_n$, the optical effective electron mass $m_n^*$, and the magnitude of
the oscillating charge $Q$. For example, one can write
$\gamma_{0n}^2=Qe/m_n^*\epsilon_2(x_n)$, \cite{Brongersma00} where $e$ is the
electron charge.
Note that $\omega_p^2=Qe/m_0^*$, \cite{Jackson1975} we therefore get
\begin{equation}
\gamma_{0n}^2=\frac{3\omega_n^2\epsilon_2(x_n)}{1+2\epsilon_2(x_n)} \equiv
\frac{1}{M_n \lambda}
\end{equation}
and
\begin{equation}
\label{eq:electronmass}
m^*_n=m^*_0\frac{[1+2\epsilon_2(x_n)]^2}{3\epsilon^2_2(x_n)}.
\end{equation}
Here $m^*_0$ denotes the optical effective electron mass of bulk metal in
vacuum.  While a free electron mass $m_0=9.1095 \times 10^{-28}$ g, typically
$m^*_0=8.7 \times 10^{-28}$ g for Ag. \cite{Johnson, Brongersma00}
Equation~\eqref{eq:electronmass} does not account for size and quantum
effects in the nanoparticle, however, it is a scale relation which indicates
that the optical effective electron mass increases from left to the right in
the GPWs. This is consistent with our terminologies of light, heavy, and
light-heavy gradons. Also it is easy to show that Eq.~\eqref{eq:bandconst} is
exactly the same as Eq.~\eqref{eq:dc_c1}.

In conclusion, we have proposed GPW structures which sustain interestingly
localized coupled plasmon modes that may have potential applications in
plasmonics, such as optical switcher and multiplexer. The underlying
localization mechanism and our analysis equally apply to a wide spectrum of
analogous problems, which may have important ramifications for understanding
excitations with transitional spectra in many condensed matter systems.

\hfill

This work was supported in part by the RGC Earmarked Grant of the Hong Kong
SAR Government (K.W.Y.), and in part by a Grant-in-Aid for Scientific
Research from Japan Society for the Promotion of Science (No.~16360044).

\newpage
\begin{center}
\textbf{Figure Captions}
\end{center}

\noindent\textbf{FIG. 1}. (Color online) Real part of the induced dipole
moment for longitudinal excitation at $i=N/2$ in a GPW for $d_0=3a$ (left
panels) and $d_0=4a$ (right panels). Damping parameter $\Gamma=0.001\omega_p$
through (a) - (f). 
Real frequencies are indicated in each panel.

\hfill

\noindent\textbf{FIG. 2}. (Color online) Dispersion relation, density of
states $D(\omega)$, and inverse participation ratio (IPR) of GPW for $d_0=3a$
(left panels) and $d_0=4a$ (right panels). Dashed-dotted and dashed lines
represent the corresponding results for PW in homogeneous host of
$\epsilon_2=3.0$  and $\epsilon_2=4.0$, respectively. (a), (d) Dispersion
relations for near field coupling. (b), (e) $D(\omega)$ versus coupled
plasmon mode frequency. The thick black lines represent results of GPW with
nearest-neighboring coupling. (c), (f) IPR versus frequency.

\hfill

\noindent\textbf{FIG. 3}. (Color online) Phase diagram showing the variation
of the four characteristic frequencies as functions of $d_0$ for (a) $c=0.5$
and (b) $c=1.0$, in the case of $\epsilon_{\ell}=3.0$. The critical
interparticle distance $d_c$ is represented by the vertical dashed lines,
whereas the two horizontal lines represent $\omega_1$ and
$\omega_{\text{N}}$, respectively.

\hfill

\noindent\textbf{FIG. 4}. (Color online) (a)--(c) Schematics of ``band
construction" in GPWs. (d) Critical interparticle distance $d_c$ as a
function of $c$ for both $\epsilon_{\ell}=3.0$ (solid line) and
$\epsilon_{\ell}=1.0$ (dashed line). The horizontal dashed line represents
$d_0=2a$, which is the geometric limit. Extended modes are possible only in
the region below these curves, e.g., shaded region in the case of
$\epsilon_{\ell}=3.0$.

\begin{thebibliography}{}
\bibitem{note0}In the near-field approximation, the results of transverse
polarizations (not shown here) are similar, except for a narrower band due to
smaller coupling strength as compared to the case of longitudinal
polarization. However, with fully retarded coupling, polariton-like
anticrossing in the dispersion is expected for transverse polarizations (see
Ref.~10). This should be true in GPWs even though phase matching is
apparently not observed.
\bibitem{note}From this argument, it is evident that the extended modes at
intermediate frequencies are not a consequence of the end effects that
constrain the high- and low-frequency response, but rather due to the
effective on-site potential ($U_n$) confinment, quite similar to Ref.~12.
Therefore, these extended modes persist even for infinite-size systems in the
infinitesimal gradient limit.
\end{thebibliography}
\end{document}